# Ridesharing in the era of Mobility as a Service (MaaS): An Activity-based Approach with Multimodality and Intermodality

Ali Najmi[a], Taha H. Rashidi[a], Wei Liu[a,b]

[a] Research Center for Integrated Transport Innovation, School of Civil and Environmental Engineering, The University of New South Wales, Sydney, Australia, 2032
[b] School of Computer Science and Engineering, University of New South Wales, Sydney, Australia, 2032
(a.najmi@unsw.edu.au, rashidi@unsw.edu.au, wei.liu@unsw.edu.au)

Mobility as a Service (MaaS), as an emerging concept, is quickly evolving and at the same time irreversibly reshaping travellers' behaviour by facilitating their accessibility to different transport modes using shared economy concepts. Managing the complex multimodal and intermodal system of quickly emerging MaaS, requires a holistic modelling framework of transport network incorporating all the influential aspects and means involved. In this paper, we mathematically formulate a novel MaaS-based activity travel pattern (ATP) generator to facilitate ridesharing in a system in which drivers and passengers interact by sharing their full activity diary. The proposed formulation extends the definition of MaaS beyond an intermodal trip planner, by incorporating an inclusive set of travel attributes including the choices of activity, activity sequence, departure time, and mode, and the transitions among the modes in ATPs of all participants. Furthermore, this paper introduces a dynamic rideshare-oriented MaaS model in which the conventional rideshare modelling structure is synchronised with the proposed MaaS-based ATP generator in a unified structure. These models explicitly bridge the gaps in the missed-out connection between ATPs planning and rideshare models in the current literature. At last, numerical examples are provided to demonstrate the significant impacts of the MaaS-based planning on the ridesharing systems performance.

## 1. Introduction

The shared mobility sector encompasses emerging ideas and platforms, which brings significant opportunities bundled with major challenges. Ride-hailing, ridesharing, and city distribution are the most well-known platforms that have already emerged. Integrating these "new" modes and other conventional modes such as public transport and providing door-to-door mobility is one of the priorities of researchers, decision-makers, transport authorities and mobility providers. Building on these shared modes and recent developments in information and communication technologies (ICT), "Mobility as a Service" (MaaS) is one of the new mobility concepts that could assist in achieving seamless mobility (Kamargianni et al., 2016). MaaS, in its common definition, combines services on different transport modes to provide customised mobility services via a single interface (MaaS Alliance, 2017). However, MaaS is in its introduction and growth stages and its implementation is still limited. Also, there exists a lack of understanding on the entire premise of MaaS and its divergence about what exactly constitutes MaaS (Wong et al., 2019). So, it would be beneficial to improve its coverage and concept beyond an intermodal journey planner.

The current MaaS platforms usually work as an intermodal journey planner which provide combinations of different transport modes such that the system participants can buy mobility services (Kamargianni et al., 2016) not a mode of transport, a membership or a subscription only. Therefore, the assumption is that the tour (a consequence of activities to be visited) is already determined and the objective is to provide the means of mobility while the mobility provision is conditional on travellers' needs. For example, activity location is a dominant factor in determining the mode of transport. If the activity is shopping, there can be different candidate locations that can be selected for this purpose. Advising the participants of a MaaS system can enhance the performance of the MaaS systems. Scheduling tour and activity visits could be a new feature to be included in MaaS systems to grow the coverage which is called a MaaS-based ATP generator in this paper. While there are some tools for integrated trip planning (e.g. TriMet in Portland) to provide multimodal trip information to participants, there is no integrated model for tour, activity visit and intermodal journey planning.

Regardless of the MaaS concept, the tour and activity schedulers have been extensively discussed in the literature. Generating trips, and choosing destinations, departure times, and modes are the core of such models where they are usually determined through some disjoint but interrelated sub-models (Najmi et al., 2018; Najmi et al., 2019a). Nonetheless, modal interactions are not explicitly considered in most traditional activity planning models mainly due to lack of spatiotemporal constraints among activity locations. Supernetwork-based models can take advantage of expanded networks to formulate multimodal TPMSs considering their interactions and spatiotemporal constraints (Najmi et al., 2019b). Still, the multimodal formulations are usually simplified in the literature so that the formulations may consider only walking, cycling, private car, and public transport at their simplest forms (Friedrich et al., 2018). Thus, despite of the prosperous future of rideshare modes in our society, its influence on travellers' ATPs is often ignored.

On the other hand, the state of the art in rideshare modelling is to formulate some matching problems with maximisation of different objective functions such as distance savings (Agatz et al., 2011), number of matches (Masoud and Jayakrishnan, 2017), and distance proximity (Najmi et al., 2017). While the performances of the ridesharing models are highly correlated with the participation rates, travel request attributes such as origins, destinations, and time windows of the requested trips, the travel attributes are usually considered fixed and sometimes stochastic but predictable in their mathematical formulations. Interestingly, the travel attributes should be the outputs of participants' ATPs but, unfortunately, the incorporation of the mode in the tour and activity schedulers and on persons' mobility has not been received enough attention neither in the literature.

In fact, the reciprocal interaction between ridesharing and MaaS-based ATP planning concepts has been neglected despite their significant mutual impacts. The main reason for this separation can be attributed to the fact that ridesharing is interpersonal and dynamic in its nature while tour planning is usually static and solely depends on individuals' choice or behavior restrictions and preferences. Nonetheless, both ridesharing and tour planning (in the form of MaaS) are platform-based which would be an incentive for their integration. Their platform-based structure discloses great opportunities in developing a rideshare oriented MaaS system for better management of travel demand by reorganising and rescheduling the travelers' activity patterns. Therefore, this paper seeks a new formulation to effectively synchronise the new MaaS-based ATP generator and rideshare concepts in a unified structure.

- ***Our contributions***

The proposed model in this paper differs significantly from cited papers in the literatures of MaaS and ridesharing modelling which is envisioned to significantly complement the existing literature. Specifically, in the current paper, we introduce a novel rideshare oriented MaaS model to be used for route planning and activity scheduling with rideshare modes at its core. Accordingly, we formulate a unified ATPs generator which incorporates spatiotemporal constraints, destination choice, multimodality, and rideshare mode. This is further expanded by providing a novel algorithm that heuristically and dynamically re-schedule the ATPs for MaaS-based participants to 1) circumvent the complexity of the model, and 2) make the model suitable to be implemented in practice. The dynamicity allows the participants in the MaaS-based planning system to adjust their tours and schedules to take more benefits from the ridesharing system by finding better matches and cheaper routes. Other than the influential effects of ridesharing in scheduling, its integration in tour planning, by itself, is a major contribution from this paper that has not been fully discussed in the literature.

We show the significant advantages of introducing ATP generator in MaaS-based systems which is armed with an ATP re-scheduler. Also, we illustrate the major contribution of MaaS-based participants in the system performance and their impacts on changing matching profile of the system. While inclusion of the participants is beneficial and improves the system-level performance measures, the contribution of conventional rideshare participants (with announcements for single trips) is decreased. This is discussed as an incentive (among others discussed in this paper) to encourage the conventional users of shared mobilities to participate in MaaS systems.

## 2. Literature review

The proposed model in this paper is built on travel scheduling, MaaS and ridesharing concepts. This section briefly reviews the background of these three pillars. Later, in Section 2.4, we further explain the research gap in the literature that has been addressed by this paper.

### 2.1 Activity scheduling and mode determination

ATP, including the determination of destination, mode, and departure time, is the main outcome of developing demand models components in activity-based models (and advanced trip-based or tour-based versions). In practice, the model usually cannot guarantee seeking the optimum ATP for each traveller as it is a computationally challenging task; instead, the models are inclined to estimate a feasible space–time region for travellers (Najmi et al., 2019b). For this purpose, using the utility maximisation-based and rule-based models is prevalent to circumvent the feasible space–time complexity. Using these models, some travelling decisions (such as trip purpose and activity sequence) are initially made for each traveller and then, using the space-time constraints for fixed activities (such as work and school), the full ATPs of travellers is scheduled based on some heuristics or rules. So, the ATPs generators in practice are not developed under unified platform which may results in asynchronisation in the system. There is a rich body of research under this category; however, we do not further illustrate the technical issues of these efforts because our focus is not on this category. Interested readers are referred to Pinjari and Bhat (2011).

In line with the efforts to develop unified ATPs generators at the presence of tempo-spatial constraints, and to model the multimodal structure of the transport systems, the integrated models, which are basically expanded network-based models, are widely applied in the literature of integrated models. In these models, to interconnect different single-modal networks, transfer links which connect the modal networks at the same physical locations were added. Following this idea, a transport model can be constructed with connecting many independent networks each of which for an individual mode-time period. The expanded network-based models may be categorised into shortest path-based and vehicle routing-based models (Najmi et al., 2019b). In the shortest path-based stream, different possible combinations of nodes and links are formed as the expanded network and the solution can be obtained by running classic shortest path algorithms onto an expanded network (Arentze and Timmermans, 2004; Liao et al., 2013; Liao, 2016; Li et al., 2018; Ramadurai and Ukkusuri, 2010). While the literature on this stream is rich, the time-dependent activity-travel assignment models are limited and not well developed (Liu et al., 2015), because the time dimension can increase the size of their network to large extent. In the vehicle routing-based stream, the optimal delivery route of travellers on the expanded network, subject to time restrictions, from a given depot to a number of predefined destinations is sought (Recker, 1995; Kang and Recker, 2013; Chow and Liu, 2012; Chow and Djavadian, 2015). These models can offer spatiotemporal constraints as the space–time prism is associated with each activity and each traveller. Continuity of time in these models results in a relatively smaller expanded network time; nonetheless, the solution algorithms for the models are much more computationally demanding than those for the time-discretised network-based models (vehicle routing-based versus shortest path algorithms).

Despite that the expanded network-based models can optimally generate the full activity patterns of travellers, their computation complexity is an obstacle that does not allow incorporating shared mobility modes such as ridesharing in their structure (formulation). Furthermore, the models are often static and their planning horizon is usually a whole day.

## 2.2 MaaS

MaaS is often described as a one-stop, travel management platform digitally unifying trip creation, purchase and delivery across all modes (Wong et al., 2019). However, MaaS is in its introduction and to some extent in growth stages so that most of its literature focus on the definitions, key elements and scope (Kamargianni et al., 2016; Jittrapirom et al., 2017).

Furthermore, there are some studies that explore various trials and case studies to identify the drivers and barriers of MaaS developments (Smith et al., 2018; Audouin and Finger, 2018). Also, different aspects of MaaS have been explored in the literature which include the potential market and demand for MaaS (Ho et al., 2018), barriers in collaboration (Smith et al., 2018), and its potential impacts on public transport (Hensher, 2017). These academic research in conjunction with many assertions in the grey literature (industry/ consultancy white papers) about the forecasted travel time savings under MaaS, and impacts on congestion and environment, have increased the passion toward MaaS research area. The interested reader on the detailed state of the art of the MaaS domain is referred to Wong et al. (2019). The authors conclude that there is a divergence in exactly what constitutes MaaS, which is a consequence of a lack of understanding on the entire premise for MaaS and the strong emotional ideology of researchers in this area.

## 2.3 Ridesharing or ride-sourcing

In ridesharing, usually a service provider matches potential drivers and passengers with similar itineraries. This allows them to travel together and share the costs. Ridesharing is dynamic in nature as the participants enter the system at any time by either requesting a drive or offering a ride (Nourinejad and Roorda, 2016). While many of the proposed models in the literature are dynamic to simulate the rideshare systems in practice (e.g. Agatz et al., 2011; Najmi et al., 2017; Xing et al., 2009), there are some models that are static mostly for policy making purposes (e.g. Amey, 2011). In ride matching models, optimisation algorithms play an important role in the ride-matching problems (Hou et al., 2018). The simplest variant of such problems only allocates to each driver a single rider, which can be formulated as a maximum-weight bipartite matching problem (e.g. Agatz et al., 2011; Najmi et al., 2017; Stiglic et al., 2016). There is a more complicated variant of rideshare models which allocate multiple passengers to each driver (e.g. Baldacci et al., 2004; Ghoseiri et al., 2011). In all the optimization variants, matching success rate has been the main concerns of scholars. Accordingly, to enhance the quality of rideshare systems, different algorithms are proposed to investigate the positive role of meeting point (Stiglic et al., 2015), matching, detour, and scheduling flexibility (Stiglic et al., 2016) on matching rates.

When determining matches between drivers and passengers, a number of constraints on the feasibility of matches must be observed. Time or spatiotemporal constraint is the common consideration, and in many cases is the only criterion, in the feasibility check. In addition to time, there are other criteria that have been used determine for feasibility considerations. The criteria include gender of participant (Levin et al., 1977; Xing et al., 2009), smoking (Ghoseiri et al., 2011) and maximum acceptable service response time (Xing et al., 2009).

Intermodal interaction is a significant component that is missed in the main body of ridesharing literature. Recently, the impact of the isolation in ridesharing has received attention in the literature. For instance the isolation may result in infeasible matchings in a ridesharing system that could be feasible if the interaction of the ridesharing and other modes are considered (Stiglic et al., 2018). Some researchers have started to test the feasibility and impacts of synchronisation between the services especially ridesharing and public transit (e.g. Mahéo et al., 2019; Stiglic et al., 2018; Yan et al., 2018). For example, Stiglic et al. (2018) examined the potential benefits of integrating ridesharing and public transport, and found that these modes are complementary. They found that public transit systems can potentially increase the use of public transit as well as the matching rate in ridesharing system. Behavioural modelling in rideshare systems is another research topic that recently have received remarkable attentions within which the travellers' participation behaviour is at the core. The models include formulation of drivers' behaviours in a ridesharing market with both electric and gasoline vehicles (Ke et al., 2019), customer confirmed-order cancellation behaviour (Wang et al., 2019) and labor supply behaviour of drivers (Sun et al., 2019).

Despite of the rich body of research on ridesharing, to the best of our knowledge, all of the mathematical models do not consider the intermodal connections and ATPs of travellers. Consider the integration of ridesharing and ATPs of travellers helps investigating both the travellers' behaviour and system-level performance in rideshare systems.

### 2.4 Research gap

MaaS definition can be extended, beyond the intermodal arrangements, to incorporate ATP generation of individuals. Accordingly, MaaS providers (platforms) play the role of tour planners and, not only control the intermodal arrangements but also determine/control/manage other travelling aspects such as destination, departure time, and visit duration. Furthermore, the ridesharing systems in the literature are designed for usually real-time applications and their optimal interactions with other modes as well as travellers' ATPs are not considered. The literature usually accounts ridesharing models isolated with predetermined demand; hence not suitable for policy appraisal targeting mode share variations. The main reason for the separation of rideshare models and ATP generators is that ridesharing is interpersonal and dynamic in its nature while tour planning is usually static and solely depends on individuals' restrictions and preferences. Therefore, other than extending the scope of MaaS, we attempt to bridge the apparent research gap in the literature between ATP generators and rideshare models where a lacuna exists to mathematically and optimally matching ridesharing participants while accounting for and improving their ATPs. All in all, assuming the principal role of MaaS-based systems for travel scheduling and redistribution of travel demand in near future, and also the prospective role of ridesharing in this mobility structure, we aim to formulate a novel rideshare-oriented MaaS system, as a tour planner, that systematically determines the optimal route and schedule of travellers. The tour planning is in a dynamic environment which is compatible with ridesharing systems. This allows us to reap the benefits of both MaaS-based and ridesharing systems.

## 3. Rideshare oriented MaaS optimisation

In this section, a static model for generating MaaS-based ATP is introduced. In Section 4, we will discuss the dynamic version of the model to be used in the MaaS provider platform. The rideshare mode plays a key role in the proposed dynamic model.

In this paper, we assume that the participants with MaaS-based requests enter the system hoping to fulfil their requirements with minimum cost. The requirements include list of activity purposes that must be visited and their time windows limitations. Also, we assume that the fixed and flexible activities and the alternative locations for conducting each of the activities are determined. Therefore, the MaaS-based ATP of a participant should reflect where the participant engages in activities, and how and when to get there. The model assumes that, for each participant in the MaaS program, the activity purposes to be visited, the location of fixed activities, candidate locations for flexible activities, and time windows for departure times and visiting durations, if any, are predetermined. The flexible destinations, departure time, mode, and visiting durations are the variables that the platform makes decision on.

Before introducing the mathematical formulation and to further clarify the problem definition, we represent the proposed network structure using the simple example in Fig. 1. In the proposed structure, nodes are either physical nodes, which are real locations in space (physical network nodes), or activity nodes (e.g., working, shopping and shared car/bicycle locations), which are never visited for moving on the physical network but represent candidate activities that may be conducted at a physical node. Further, any link is either a travel link, on which a movement can happen on a physical network (e.g., private vehicle road, path foot, ridesharing), or a virtual link, which never changes the location of a traveller (here the participant in MaaS-based ATP planning) but allows a transition between different modes or an activity conduction (e.g., links adjacent to activity nodes).

Each MaaS-based participant can have a flexible role meaning that he/she can be flexible to perform either driver or passenger roles. Depending on the spatiotemporal constraints, the platform decides on the roles and establishing rideshares over time. An example of ATPs for participants P1 and P2 are shown in Fig. 1. Participant P1 leaves home (by private car) to conduct an activity at node 13 for about 4.5 hours; however, his/her itinerary has slightly coincide with the potential route of participant P2 who intends to conduct two activities at nodes 12 and 14. Thus, participant P1 drives to node 6 to pick up participant 2 who has walked to this node. Participant P1 drops off participant P2 at node 14 and then continues his/her route to his/her destination (activity location A1). Participant 2 leaves activity A2 at 10:35 toward node 12 to conduct activity A3. For this purpose, he/she uses a carsharing mode and uses the shared car at node 14. After conducting activity A3, he/she returns the shared car to node 12 where it had been previously parked. To return home, participant P2 arranges another ride with participant P1 at node 14 as the meeting point. Participant P1 drops off participant P2 at node 6 and drives to his/her home H1. Finally, participant P2 walks from node 6 to node 2, where his/her home is located. It should be noted that not all the possible modes (e.g. bicycle and public transport) and links (e.g. all the ridesharing links) are represented in this figure.

**Figure 1** expanded network representation

The problem of visiting a set of activities on a network is a vehicle routing problem (VRP) which forms the main basis of our ATP formulation. Thus the problem is NP-hard in complexity (Lenstra and Rinnooy Kan, 1981), and sensitive to the number of the nodes in the network. Thus, to simplify the problem, we propose a pre-processing step, to prepare and scale down the solution space, and a post-processing step to map the solution to the physical network. The pre-processing, post-processing and ATPs formulation are discussed in the following paragraphs.

### 3.1. Pre-processing

Optimising the original connected networks (as in Fig. 1) is quite challenging because the large numbers of nodes and links within the network represent a substantial computational challenge in solving the problem. To circumvent this obstacle, the original connected networks should be mapped into the activity level networks which include much fewer number of nodes. However, each participant should have its own activity level network which includes only the candidate destinations that might be visited within his/her ATP. For MaaS planning purposes, we assume that the participants provide their candidate destinations for conducting their activity to platform in advance. Still, the platform may add some additional nodes called rideshare meeting points, if necessary. At the activity level, the cost of each link corresponds to the shortest path on the original connected network. Nonetheless, this reduction does not affect the solution space as the ATP on the original expanded network could be readily retrieved in a post-processing step by mapping the activity layer to the physical network. In mapping the physical network to the activity level network, the represented links for different modes should be distinguished.

Fig. 2 shows how the physical network of Fig. 1 can be converted into an aggregated level for participant P2. In the figure, as the meeting point node is an effective node in the solution space, the node is included in the aggregated level. The participant can leave home either by requesting a ride or on foot. He/she can choose between activities A2 and A3 to go first. If the participant goes to activity A3, he/she must go to activity A2 on foot. On the other hand, if the participant chooses activity A2, he/she can choose between carsharing and walking modes to get to activity A3. If the participant uses the shared car, he/she must return it to the same location. Thus, pre-processing allows deducing the size of the network significantly.

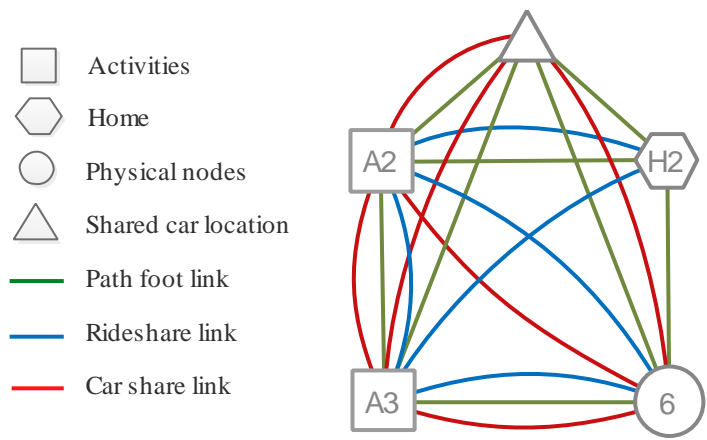

**Figure 2** Aggregated level for traveler P2 in Pre-processing

### 3.2. Activity travel pattern optimisation

To achieve the optimum ATPs for the participants, in this section, a mixed integer linear programming problem is presented in which visiting activities in the space-time prism with the least cost is the core in determining the ATP for each passenger.

Consider an activity level network $N(V,E)$ composed of a set of nodes, $V$, and a set of directed links, $E$. Let $(i,j) \in E$ denotes a directed link in $N$ which connects node $i$ to $j$ and allows the participant $p \in P$ to traverse the links to conduct some activities. We assume that all possible daily activity-travel patterns should be conducted over the time period [0,T]. Each node is associated with a departure time window constraint $[\underline{t}_i^p, \bar{t}_i^p]$, where $\underline{t}_i^p$ and $\bar{t}_i^p$ are the earliest and latest time to departure from node $i \in V$ by participant $p \in P$; and an activity duration $[\underline{d}_i^p, \bar{d}_i^p]$ constraint, where $\underline{d}_i^p$ and $\bar{d}_i^p$ are the minimum and maximum durations that participant $p \in P$ can spend at node $i \in V$. The departure may happen by different modes of transport $k \in K$. We denote link selection variable $x_{ij}^{p,k}$, which takes the value of 1 if node $j \in V$ is visited immediately after visiting node $i \in V$ by participant $p \in P$ and mode $k \in K$, and 0 otherwise. The variable determines the type and the sequence of visits to the nodes.

Parameters

| | |
|---|---|
| $N(V,E,p)$ | Network for participant $p$ |
| $p \in P$ | participant $p$ |
| $D^S, R^S \subset P^S$ | subset of single-trip based participants representing drivers and passengers |
| $D^M, R^M \subset P^M$ | subset of MaaS-based based participants representing drivers and passengers |
| $i \in V$ | nodes in the network |
| $k \in K$ | mode of transport |
| $C \subset V$ | subset of nodes representing shared car/bicycle locations |
| $i \in V(SC)$ | shared car/bicycle nodes in the network |
| $\xi_p^*$ | route cost for participant $p$ without detour |
| $(i,j) \in B$ | links in the subtour |
| $M$ | big constant |
| $(i,j) \in E(k)$ | links of type $k \in K$ in the network which are categorised into road links $A$ (for private/shared car or bicycle), ridesharing $RS$, and other links $O$ (for walk and public transpirt) |
| $RS$ | rideshare mode |
| $\bar{\tau}_{ij}$ | distance on link $(i,j) \in E$ |
| $\tau_{ij}$ | travel time on link $(i,j) \in E$ |
| $[\underline{d}_i^p, \bar{d}_i^p]$ | limits of the spent time on node $i$ by participant $p \in P$ |
| $[\underline{t}_i^p, \bar{t}_i^p]$ | limits of the departure time from node $i$ by participant $p \in P$ |
| $h$ | time step in rolling horizon algorithm |
| $T$ | set of time step in rolling horizon algorithm |
| $e(p)$ | earliest departure time of single-trip participant from his/her origin |
| $l(p)$ | latest arrival time of single-trip participant to his/her destination |
| $\delta \in E(p)$ | link belong to edges for participant $p$ |

Variables:

| | |
|---|---|
| $x_{ij}^{p,k}$ | 1 if node $j \in V$ is visited by mode $k \in K$ immediately after visiting node $i \in V$ by participant $p \in P$ and 0 otherwise |
| $t_i^p$ | departure time from node $i \in V$ by participant $p \in P$ |
| $d_i^p$ | time spent at node $i \in V$ by participant $p \in P$ |
| $y_{ij}^{p,p'}$ | 1 if participants $p, p' \in P$ are matched on link $(i,j) \in E(R)$ and 0 otherwise |

There are several behavioural restricting assumptions that should be considered in the model formulation. An ATP is valid if it fulfils all the following rules:
1- A participant can leave home with a private car, bicycle, public transport, sharing a ride, or on foot to conduct out-of-home activities.

2- If a participant leaves home by a private car/bicycle, he/she needs to return to his/her home with the same private car/bicycle at the end of their tour.
3- If a participant collects a shared car/bicycle for a trip, he/she must return the car to the same location that had been collected from, before the end of the tour.
4- If a participant leaves (parks) his/her private vehicle at an activity location, he/she must return to the place to remove the car.
5- A participant cannot switch to the private vehicle network if he/she has left his/her home without a private vehicle.

From here on, the model constraints are discussed. Constraint (1) is the flow conservation constraint and indicates that, in an ATP, if a node is visited, it must also be left.

$$\sum_{i:(i,j)\in E(k)} x_{ij}^{p,k} - \sum_{i:(j,i)\in E(k)} x_{ji}^{p,k} = 0 \qquad \forall p \in P^M, \ \forall j \in V \qquad (1)$$

There are some nodes in the system that are visited several times. For example, if a shared car/bicycle is picked up by a participant, it must be returned to the same location before the end of the trip. Therefore, flow conservation constraints are needed for private/shared vehicles in which ensures that if a vehicle (private or shared) get to a node, it must leave it. It does not mean that if a participant chooses a private vehicle to leave home with, he/she must use the same mode throughout the route. Participants may leave (park) their private vehicle at an activity node and change their mode; however, they must return to the same node (multi-visit) to remove the private/shared car/bicycle from the node. It should be mentioned that implementing this multi-visit possibility needs to be addressed in pre-processing by either enumerating the nodes or links (Berghman et al., 2014; Kinable et al., 2014) or adding visits possibilities in to the formulation (Najmi et al., 2019b). Enumerating the nodes is implemented only for shared car nodes (as in Constraint 3) in this paper which can be extended to other nodes.

$$\sum_{j:(i,j)\in E} x_{ij^+}^{p,k} - \sum_{j:(i,j,)\in E} x_{ij^-}^{p,k} = 0 \qquad \forall p \in P^M, \qquad \forall k \in K^*, \forall j \in V(CS) \ \exists (j^+, j^-) \qquad (2)$$

Constraint (3) is a time-window constraint and ensures that a feasible route (sequence of nodes) in the space-time prism will be selected. We denote the travel time on the link $(i,j)$ by $\tau_{ij}$.

$$t_i^p + \tau_{ij} + d_j^p - t_j^p - M\left(1 - x_{ij}^{p,k}\right) \leq 0 \qquad \forall p \in P^M, \forall k \in K, \forall (i,j) \in E(k) \qquad (3)$$

Constraint (4) is the connectivity constraint (sub-tour elimination).

$$\sum_{\substack{k\in K, (i,j)\in E(k) \\ i\in B, j\notin B}} x_{ij}^{p,k} \geq 1 \qquad \forall p \in P^M, \ \forall B \subset V \qquad (4)$$

where $B$ is a sub-tour formed in the ATP solution. The provided connectivity constraint is an extension of the traditional sub-tour elimination constraint originally developed by Dantzig et al. (1954). In every ATP solution, the constraint forces at least one edge pointing from $B$ to its complement. This means $B$ cannot be disconnected. In this constraint, every node $i \in B$ must be the origin of one edge to another node of $j \in B$ or to a node $j \notin B$.

We assume that each participant has a set of activity types that to be conducted within the spatiotemporal constraints. Some activity types, such as school and work may have only a single candidate destination, whereas each flexible activity such as shopping can be conducted in one of multiple candidate locations. Therefore, no more than one of the candidate locations can be visited. Therefore, the problem is under the category of generalised VRP (Laporte and Nobert, 1983; Noon and Bean, 1991) in which the node set $V$ is partitioned into several clusters $C^p$ and the goal is to determine the best cycle, starting and ending at home, and visiting no more than one node in each cluster (see Constraints (5)). Different candidate locations for each activity, assuming the same activity duration, offer the same satisfaction.

$$\sum_{\substack{i \in \psi, k \in K \\ j:(i,j) \in E(k)}} x_{ij}^{p,k} = 1 \qquad \forall\, p \in P^M, \forall \psi \in C^p \tag{5}$$

Constraints (6) and (7) ensure that departure times and activity durations are properly handled. It should be noted that the time window for departure time and minimum duration of the non-activity nodes are $[0, T]$ and 0, respectively.

$$\underline{t}_i^p \leq t_i^p \leq \bar{t}_i^p \qquad \forall\, p \in P^M, \forall\, i \in V \tag{6}$$

$$\underline{d}_i^p \leq d_i^p \leq \bar{d}_i^p \qquad \forall\, p \in P^M, \forall\, i \in V \tag{7}$$

Constraints (8) and (9) specify the range and values of the decision variables.

$$x_{ij}^{p,k} \in \{0,1\} \qquad \forall\, p \in P^M, \forall\, k \in K, \forall\, (i,j) \in E(k) \tag{8}$$

$$t_i^p, d_i^p \in \mathbb{R}_{\geq 0} \qquad \forall\, p \in P^M, \forall\, i \in V \tag{9}$$

Having a properly pre-processed transport network (as explained in section 3.1), the modes of walking, private/shared car and bicycle, and public transport (with some simplifying assumptions) can be addressed by the equations discussed above. However, addressing the rideshare mode in the multimodal structure is complicated because of its interpersonal nature which considers the interaction of participants in the system. The rideshare mode is formulated in the next section.

*Rideshare modelling*

As it was mentioned before, each of the participants has its own customised activity-level network. Therefore, the nodes of different participants usually are different. Nonetheless, to incorporate the rideshare mode in the formulation, the pre-processing stage should be extended. For each participant, their activity-level network should be extended to incorporate the other participants' feasible links (and their corresponding nodes) for ridesharing. The feasibility conditions are fully discussed in Section 4.3 where the focus is on feasibility check of participants in ridesharing framework. The network customisation is explained using the example of Fig. 3. Suppose giving a ride on link *L2* (which is for another participant) is feasible for the driver if the driver has a detour to give this ride instead of traversing its *L1* link. So, all nodes and links related to this detour should be added to the activity-level network of the driver.

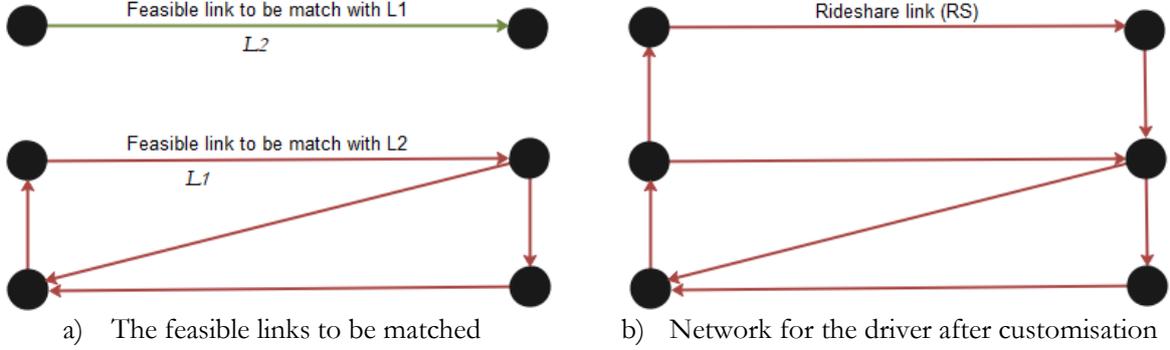

**Figure 3** Customising the network for a driver to incorporate the possibility of giving a ride to another participant over link L2

Having the customised networks and tagging the rideshare links (RS), the ATP formulation can be extended to incorporate the rideshare mode. Constraint (10) ensures that if participants $p \in D$ and $p' \in R$ are matched on the link $(i,j) \in E(R)$, both participants must traverse the link. In the formulation, $y_{ij}^{p,p'}$ represents the matching variable which is 1 if driver $p \in P$ and passenger $p' \in P$ are matched on link $(i,j) \in E(RS)$ and 0 otherwise.

$$x_{ij}^{p,A} x_{ij}^{p',RS} - y_{ij}^{p,p'} = 0 \qquad \forall\, p \in D^M, \forall\, p' \in R^M, \forall\, (i,j) \in E(RS) \qquad (10)$$

Constraint (11) enforces that a participant $p' \in R$ can traverse a rideshare link only if it is matched with one driver.

$$x_{ij}^{p',RS} - \sum_{p \in D} y_{ij}^{p,p'} = 0 \qquad \forall\, p' \in R^M, \forall\, (i,j) \in E(RS) \qquad (11)$$

Constraint (12) specifies the departure time of the driver and passenger to be the same if they are matched.

$$\sum_{l \in L} y_{ij}^{p,p'} \left( t_i^{p'} - t_i^{p} \right) = 0 \qquad \forall\, p \in D^M, \forall\, p' \in R^M, \forall\, (i,j) \in E(RS) \qquad (12)$$

Constraints (13)-(14) are matching constraints and guarantee that, in each of the links $(i,j) \in E(RS)$, each passenger is matched by at most one driver and each driver is matched by at most its available seats, $Cap^p$. In other words, it is a car-pooling formulation which is incorporated in the model. In the cases where $Cap^p = 1$, the formulation reduces to the peer-to-peer ridesharing formulation.

$$\sum_{p' \in P} y_{ij}^{p,p'} \leq Cap^p \qquad \forall\, p \in D^M, \forall (i,j) \in E(RS) \qquad (13)$$

$$\sum_{p \in P} y_{ij}^{p,p'} \leq 1 \qquad \forall\, p' \in R^M, \forall (i,j) \in E(RS) \tag{14}$$

**Objective function**

A weighted objective function is defined in Eq. (15).

$$\begin{aligned}
\mathbf{min} \quad & \sum_{\substack{k \in K, (i,j) \in E(k) \\ p \in P^M}} \bar{\tau}_{ij} x_{ij}^{p,k} + \sum_{\substack{k \in K, (i,j) \in E(k) \\ p \in P^M}} \left(t_j^p - t_i^p - d_j^p - \tau_{ij}\right) x_{ij}^{p,k} \\
& - \sum_{\substack{k \in K, (i,j) \in E(k) \\ p \in R^M}} \bar{\tau}_{ij} x_{ij}^{p,RS}
\end{aligned} \tag{15}$$

The first term of the objective function is the total travel time over the transport system. This term is equivalent to the total travel distance over the system under the assumption that travel time is a linear function of travel time (in the case study of this paper, we have used only the minimisation of total travel distances for analysis). The second term takes into account the waiting time over all nodes. A portion of time spent at the nodes is subject to waiting time. The ideal value for waiting time is 0; nonetheless, the time windows on departure times of the activity nodes, time windows on the activity durations, and the waiting time at the meeting point for rideshare participants may result in undesirable waiting times. As the travelled distance over each of rideshare links are considered twice in the first term of the objective function, once for driver and once for passenger, one of them should be excluded. Therefore, the third term of the objective function refers to the deduction of the distances travelled over the rideshare links by passengers.

### 3.3. Post-processing

In the post-processing, after finding the ATPs in activity level, the generated ATPs for all the participants are mapped to the physical network to obtain the exact route of the participants. While the sequence of activities and the shortest path among each pair of nodes are available, the physical routes can be obtained.

### 3.4. Complexity

The developed formulation in this paper accommodates the complex interactions between different modes, including the ridesharing, and their interconnections with the participants' decisions choices. Excluding the rideshare mode, the model is a VRP. Finding a feasible solution for a VRP with time window problem in itself is an NP-complete problem (Savelsbergh, 1985); nonetheless, solving a VRP with a reasonable number of nodes, which is the case in activity routing problem as people usually visit a small number of activity type over a day, is not a problem. The rideshare mode which encompass the interaction between different participants in conjunction with the generalisation version of VRP with time window makes solving the model computationally cumbersome. To illustrate further, solving the model for a tiny example of a handful participants take about a long time which indicates the complexity of the formulation. Therefore, we propose a dynamic version of the model in Section 4 to heuristically solve the problem in large-scale cases.

## 4. Dynamic rideshare oriented MaaS optimisation

The proposed ridesharing constraints of the rideshare oriented MaaS model (eq. (10)-(14)) significantly increase the complexity of the model. To overcome this difficulty, in this section, we introduce the dynamic version of the model in which the system periodically solves a number of smaller problems and re-optimise the ATPs of participants in a stagewise manner, based on the active rideshare opportunities/offers in the solution space. In this regard, we first explain an additional pre-processing to find the feasible matches and then introduce a rolling horizon algorithm to solve the problem. Furthermore, to generalise the algorithm to rideshare models and to increase the quality of the solutions in the MssS-based ATP generator, we also incorporate participants with single-trip announcements in the system. Obviously, the system collapses to a conventional rideshare model if the MaaS-based participants are replaced with participants with single-trip announcements. In the next section and before explaining the dynamic rideshare oriented MaaS model structure, we discuss the participants with single-trip announcements and their attributes in detail.

### 4.1. Participants with single-trip announcements

Participants with single-trip announcements are participants who either request a single ride or offer a single drive while their origin and destinations are predetermined and fixed. The single-trip announcements are equivalent to the announcements in the conventional rideshare models which are looking for matches to fulfil their single trips. The presence of announcements in the system has a remarkable impact on the performance of the whole system as not only they increase the success rate of finding a match in the system but also do increase the attractiveness of the MaaS mode for the participants.

Consider a set of single-trip based drivers $D^S$ and a set of riders $R^S$. We denote $i(p)$ and $j(p)$ the origin and destination of a single-trip participant $p \in D^S \cup D^S$. For each participant $p$, the earliest time by when the participant can depart from his/her origin $\underline{t}_i^p$, and its latest arrival time at his/her destination $l(p)$ are assumed to be given. Furthermore, the pairwise distances between all node locations $\bar{\tau}_{i(p)j(p)}$ are assumed to be known. In this paper, we assume that the travel cost is a linear function of travel distance. The departure time window of announcement $p$ is then $[\underline{t}_i^p, q(p)]$ where $q(p) = l(p) - \tau_{i(p)j(p)}$ is the latest departure time of the participant. In the ridesharing literature, the length of the departure time window, $f(p) = q(p) - \underline{t}_i^p$, is known as the *flexibility* of participants (Agatz et al., 2011). It should be noted that we have tried to keep the consistency of the notations with those for the MaaS-based formulation.

## 4.2. Individual route optimisation

A MaaS-based participant uses the rideshare mode if his/her participation results in lower cost. Therefore, we call a rideshare participation *feasible* if it results in cost reduction. To find feasible matches, detour lengths before and after detour should be calculated to find out if the detour results in lower cost or not. To calculate the cost before detour, the model in Section 3.2 should be optimised for each of the MaaS-based participants in the system to find out their individual optimum routes cost. The route includes their category of participant (driver or passenger), and the modes that they will use to get their activity locations. Therefore, let denote $\xi_p^*$ the individual route cost for participant $p$. The current best route plays key roles in rideshare system where the MaaS based participants will be asked to modify their route if they are matched to a passenger/driver so that a more efficient route (with lower cost) is obtained. The application of $\xi_p^*$ will be discussed in the next section where we are looking for the feasible matches for MaaS-based participants. The calculations of tour and detour lengths are explained in detail in the next section.

## 4.3. Matching feasibility

In this section, the matching feasibility of each pairs of participants is investigated. It should be mentioned that, in the ridesharing literature, the passengers are commonly called "rider" as they exclusively enter the system to receive a ride for a single trip; nonetheless, in MaaS, the passengers are planned for ATPs which are usually using different modes. For consistency, we use the "passenger" term for all participants, including riders, who do not use private vehicles. Furthermore, for MaaS-based participants, depending on the mode used from home, the participant can be categorised into drivers and passengers. Accordingly, the passengers cannot change their roles until they return home. This is not the case for drivers as they may leave their cars somewhere en-route and change their role to passenger. The dynamic model can cover changing the roles provided that the home-based departure mode is taken into account in the pre-processing of all the iterations of the model.

Having the driver/passenger and single-trip/MaaS-based categories of announcements, the participants in the system are categorised into four groups of 1) single-tip drivers, 2) single-trip passengers, 3) MaaS-based drivers, and 4) MaaS-based passengers. Therefore, each participant $p$, whether his/her request is for a single-trip or MaaS ATP, is either a driver ($D^S \cup D^M$) or a passenger ($R^S \cup R^M$) looking for their best matches using the platform. In the first two categories, the request is for one trip while in the last two categories, the request is for a tour. We, henceforth, refer to $d$ and $r$ participants as the driver role and the passenger role, respectively. Furthermore, we assume that each shared ride consists of a single pick-up and a single drop-off.

The detour by driver $d$ to give a ride to passenger $r$, henceforth, is referred to as the pair $(d, r)$, which may be *feasible* if 1) the time windows of driver $d$ and passenger $r$ overlap, and 2) the detour costs of the participants are justifiable. According to the second feasibility check, we consider a match feasible if it results in a cost reduction (saving). For the sake of brevity, we assume that the cost is a function of travelled distance; so, the feasible match should result in distance saving. Depending on the type of participants' requests, the matching pairs $(d,r)$ can be categorised into: 1) single trip drive announcement and single trip ride announcement (S&S), 2) MaaS-based driver and single-trip passenger (M&S), 3) single-trip driver and MaaS-based passenger (S&M), and 4) MaaS-based driver and MaaS-based passenger (M&M). Following, we discuss procedures to check the feasibility of the pairs under these categories.

- *S&S feasibility check*

In single-trip cases, there is only one announced trip (link) for the participants. To check for a S&S feasibility we can calculate the overlap between the driver and the passenger time windows by comparing the latest time by when the driver must depart $k_d = \min[\, l(r) - \tau_{i(r)j(r)} - \tau_{i(d)i(r)}\,,\ l(d) - \tau_{j(r)j(d)} - \tau_{i(r)j(r)} - \tau_{i(d)i(r)}\,]$ and the earliest departure times of both the driver and the passenger where $l(d)$ and $l(r)$ are the latest arrival time at the driver and passenger destinations, respectively. The match $(d, r)$ is time feasible ($TF(d,r) = \text{True}$) only if $\Delta T_d = k_d - Max[t, \underline{t}_i^d] \geq 0$ and if $\Delta T_r = k_d + \tau_{i(d)i(r)} - Max[t, \underline{t}_i^r] \geq 0$, where $t$ is the execution time (will be explained in Section 4.6) at which the problems, including the updated ATPs and matching problems, are solved.

According to the second feasibility check, a detour cost/distance is feasible if it results in a cost/distance saving. Distance saving is a common matching criterion that has been widely used in the ridesharing literature. The interested reader is referred to Najmi et al. (2017) and Agatz et al. (2011) for more details. In a detour by a driver to give a ride to a passenger, the driver does not drive the original trip, instead he/she select the new route to pick up the passenger at $i(r)$ and drive the distance $\bar{\tau}_{i(r)j(r)}$ to the passenger's destination $j(r)$ and then continue his/her tour. A detour results in distance savings only if the length of the matched trip – including the pick-up trip, the shared trip and the drop-off trip – $S_u(d,r) = \bar{\tau}_{i(d)i(r)} + \bar{\tau}_{i(r)j(r)} + \bar{\tau}_{j(r)j(d)}$ is shorter than the sum of the lengths of the individual respective trips for both the driver and the passenger $S_v(d,r) = \bar{\tau}_{i(d)j(d)} + \bar{\tau}_{i(r)j(r)}$ (Najmi et al., 2017). Therefore, the net distance savings is $\Delta S(d,r) = S_v(d,r) - S_u(d,r)$. Fig. 4 elaborates the concept.

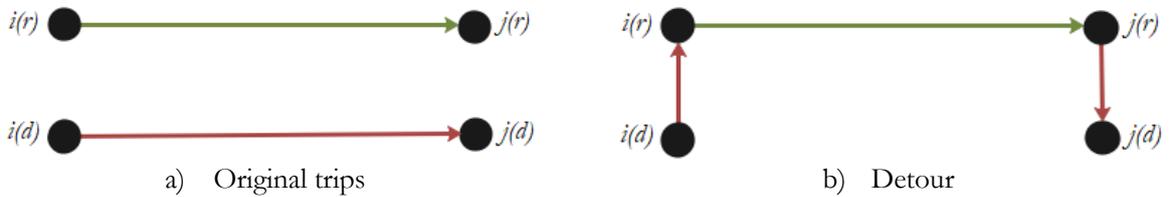

a) Original trips          b) Detour

Fig. 4 Detour feasibility of pair $(d,r)$ in S&S

- *M&S feasibility check*

In the cases of MaaS-based driver and single-trip passenger, we assume that the driver changes his/her route or ATP to be able to give a ride to a passenger. In other words, the driver does not perform its original route instead he/she select a new route or totally change his/her ATP give a ride to a passenger. Therefore, in the new route, the origin $i(r)$ and destination $j(r)$ of the passenger are the nodes that must be visited while the equations (1)-(9) are satisfied. Therefore, the match $(d, r)$ is time feasible ($TF(d,r) = \text{True}$) only if there is a feasible solution for the routing problem given in section 3.2 that traverse link $r$ (by enforcing $x_r^{r,RS} = 1$.). Fig. 5 shows an example of the route length for M&S cases.

To check the distance feasibility, the length of the ATP after matching $S_u(d,r)$ would be equal to $\sum_{(i,j)\in E(A)} \bar{\tau}_{ij} x_{ij}^{d,A}$. Furthermore, the sum of the length of optimum individual route cost $\xi_d^*$ for driver $d$ (if the driver does not give a ride at all) and the length of the individual trip for passenger $r$ forms the $S_v(d,r)$. Then, the net distance savings $\Delta S(d,r)$ can be obtained in the same formulation as for S&S.

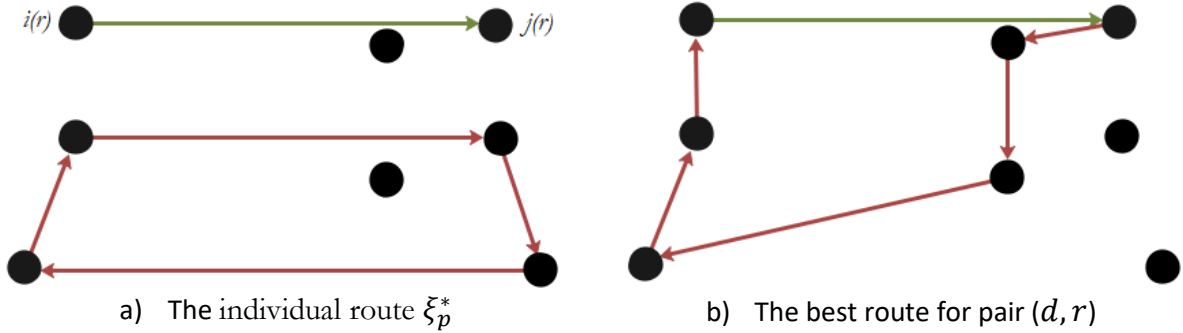

a) The individual route $\xi_p^*$  b) The best route for pair $(d, r)$

Figure 5 Detour feasibility of pair $(d, r)$ in M&S

- ***S&M feasibility check***

In the cases of single-trip based driver and MaaS-based passenger, we assume that both driver and passenger can change their original trips and routes so that the driver would be able to give a ride to the passenger. Therefore, this is different from the S&S and M&S cases because the rideshare link on which the driver and passenger will be matched is unknown. There might be multiple potential links on which a pair $(d,r)$ can be matched. Henceforth, we recognise the links as *candidate rideshare links*. Accordingly, let extend the definition of matching pair by including the candidate rideshare link. Therefore, we denote $(d,r,\delta)$ the detours by driver $d$ and passenger $r$ to, respectively, give and receive a ride on link $\delta$. In a similar manner, the definitions of $S_u$, $\Delta S$ and $TF$ are extended. An example of the S&M cases is shown in Fig. 6. Matching feasibility should be performed for each candidate rideshare links $\delta$ in the activity-level network of passenger $r$ to find out the possibility and quality of matching driver $d$ and passenger $r$.

Calculating the time feasibility for S&M cases are the same as for M&S cases; however, the time limits of driver should be applied for the origin and destination of $\delta$ before running the ATP generator for passenger $r$. Accordingly, the $\underline{t}_{i(\delta)}^r$ and $\bar{t}_{j(\delta)}^r$ should be replaced with $\max\{\underline{t}_{i(\delta)}^r, e(d) + \tau_{i(d)i(\delta)}\}$ and $\min\{\bar{t}_{j(\delta)}^r, l(d) - \tau_{j(\delta)j(d)} + \underline{d}_{j(\delta)}^r\}$, respectively. Therefore, to check the time feasibility for pair $(d, r, \delta)$, the ATP generator should be solved for passenger $r$ by enforcing $x_\delta^{r,RS} = 1$. In the case of existing a feasible solution, the match $(d,r, \delta)$ is time feasible ($TF(d,r,\delta) = \text{True}$). The sum of the lengths of the tours after matching driver $d$ and passenger $r$ over link $\delta$ is equal to $S_u(d,r,\delta) = \sum_{k\in K,(i,j)\in E(K)} \bar{\tau}_{ij} x_{ij}^{r,k} + \bar{\tau}_{i(d)i(\delta)} + \bar{\tau}_{j(\delta)j(d)}$ where $i(\delta)$ and $j(\delta)$ are the origin and destination of link $\delta$ that is used to match driver $d$ and passenger $r$. Also, the sum of the length of the individual route $\xi_r^*$ for passenger $r$ (if it is not matched with driver $d$) and the length of the individual trip for driver $d$ $-S_v(d,r)$ are calculated. Similar to M&S, the net distance savings $\Delta S(d,r,\delta)$ can be obtained; the match is space feasible only if $\Delta S(d,r) \geq 0$.

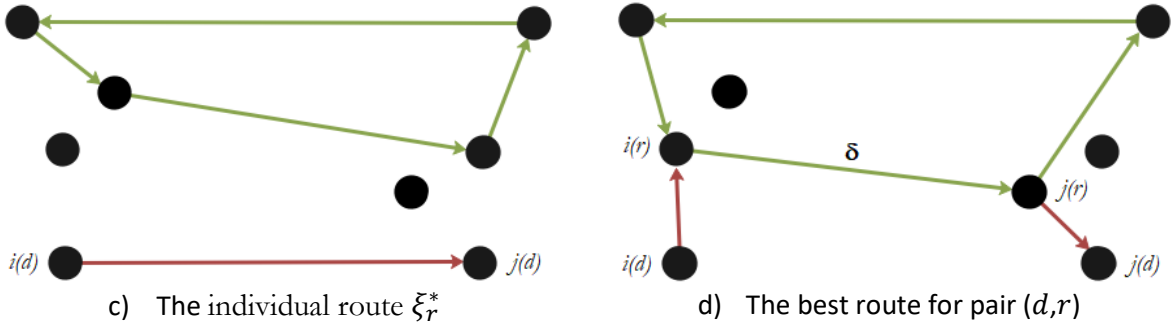

c) The individual route $\xi_r^*$ 　　　　d) The best route for pair $(d,r)$

Figure 6 Detour feasibility of pair $(d,r)$ in S&M

- *M&M feasibility check*

An example of the M&M case is depicted in Fig. 7 in which both driver and passenger may have a detour. Solving the formulation provided for ridesharing in section 3.2, for only one driver and one passenger is not complicated. So, to check the time feasibility for pair $(d,r,\delta)$ and to calculate the total length of distances travelled by the participants after matching, the model should be solved by enforcing $x_\delta^{r,RS} = x_\delta^{d,A} = 1$. In the case of existing a feasible solution, the match $(d,r,\delta)$ is time feasible ($TF(d,r,\delta) = \text{True}$). To check its space feasibility, like the other variants, the net distance savings $\Delta S(d,r,\delta)$ should be calculated. The sum of the lengths of the tours after matching $S_u(d,r,\delta)$ is equal to $\sum_{(i,j)\in E(A)} \bar{\tau}_{ij} x_{ij}^{d,A} + \sum_{k\in K,(i,j)\in E(K)} \bar{\tau}_{ij} x_{ij}^{r,k} - \bar{\tau}_\delta$. Also, the sum of the length of the current best route for driver $d$ and passenger $r$, before matching, is $\xi_d^* + \xi_r^*$ which is equivalent to current individual best route lengths $S_v(d,r)$.

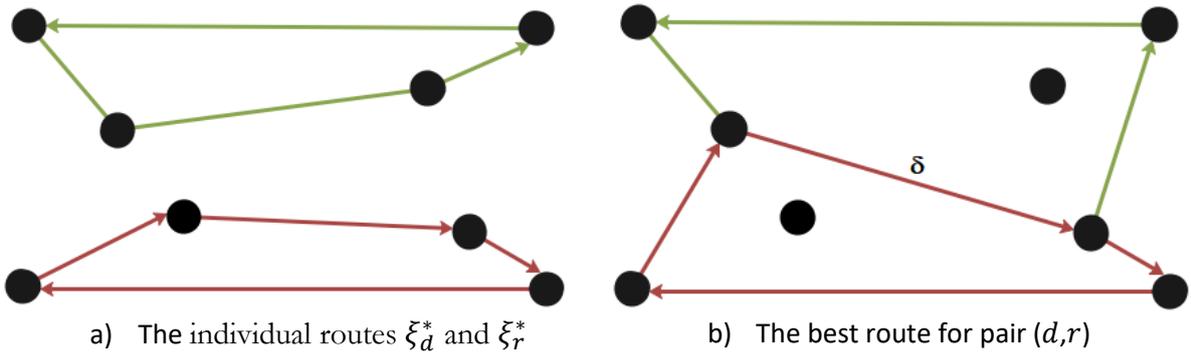

a) The individual routes $\xi_d^*$ and $\xi_r^*$ 　　　　b) The best route for pair $(d,r)$

Figure 7 Detour feasibility of pair $(d,r,\delta)$ in M&M

If we denote $P = \{(d,r,\delta): d\epsilon D, r\epsilon R, \delta\epsilon E(d) \cup E(r)\}$ the set of all pairs of drive-passenger-common links. According to the explanations in the aforementioned paragraphs, the subset of feasible pairs is identified. Therefore, let denote $\bar{P}$ be the set of feasible pairs: $\bar{P} = \{(d,r,\delta) \in P: d\epsilon D, r\epsilon R, \delta\epsilon E(d) \cup E(r), \Delta S(d,r,\delta) \geq 0, TF(d,r,\delta) = \text{True}\}$. The $\bar{P}$ set is used in the matching problem.

### 4.4. Matching drivers and passengers

In sections 4.3, the distance saving of each feasible pair $(d, r, \delta)$ for matching is calculated. In this section, we present a matching algorithm to optimally determine which drivers and passengers and how they should be matched to each other. Let use the notation of $(d, r, \delta)$ for all the matching variants to keep the consistency of the formulation; obviously, $\delta$ refers to at most one link in S&S ad M&S variants. Also, we represent the driver-passenger interactions as a bipartite graph $G$, such that each pair of participants and the corresponding link $(p, \delta)$ is represented by a node and the nodes are classified into two major sets of drivers-rideshare links and passengers-rideshare links. Formulating a matching problem over these two major sets is equivalent to a bipartite graph matching problem. Bipartite graphs are a particular class of graphs whose nodes can be divided into two disjoint sets, in which only the link between two nodes in different sets is permitted (Guillaume and Latapy, 2006; Blattner et al., 2007). However, the problem in hand is to some extent different. In our problem, each of the major sets are partitioned into $|D^S \cup D^M|$ and $|R^S \cup R^M|$ minor groups, respectively, where at most one link from/to each partition is permitted. Therefore, we provide a generalised version of bipartite graph matching problem in which the partitioned groups are incorporated.

Let $z_{d,r,\delta}$ be a binary decision variable equal to 1 if the pair $(d, r, \delta)$ is matched, and 0 otherwise, and let $w_{d,r,\delta} \geq 0$ be a weight representing the contribution of matching driver $d$ and passenger $r$ respectively over their $l$ and $l'$ links. The objective of the ridesharing matching problem is to maximise the weighted sum $\sum_{(d,r,\delta) \in \bar{P}} w_{d,r,\delta} z_{d,r,\delta}$ and the complete formulation of the problem is summarised in Eq. (1)-(4).

$$\max \sum_{(d,r,\delta) \in \bar{P}} w_{d,r,\delta} z_{d,r,\delta} \tag{16}$$

*Subject to:*

$$\sum_{\substack{r \in R, \delta \in E(d): \\ (d,r,\delta) \in \bar{P}}} z_{d,r,\delta} \leq 1 \qquad \forall d \in D^S \cup D^M \tag{17}$$

$$\sum_{\substack{d \in D, \delta \in E(r): \\ (d,r,\delta) \in \bar{P}}} z_{d,r,\delta} \leq 1 \qquad \forall r \in R^S \cup R^M \tag{18}$$

$$z_{d,r,\delta} \in \{0,1\} \qquad \forall (d,r,\delta) \in \bar{P} \tag{19}$$

The weights $w_{d,r,\delta}$ of each edge play a critical role in forming the best solution. We use the index of distance saving $\Delta S(d, r, \delta)$, discussed in Section 4.2, for each pair $(d, r, \delta) \in \bar{P}$ in the objective function for numerical experiments in Section 5. Other indexes can also be used for $w_{d,r,\delta}$. For example, assigning the same weight to all the pairs may be the best option for the problems that are not dynamic and the objective is to obtain the maximum number of matches (NM); nonetheless, assigning the same weight may not be appropriate for many dynamic and real-time problems.

### 4.5. "STATIC" solution algorithm

The steps that are illustrated in the previous Sections are summarised in Algorithm 1, henceforth referred to as STATIC. This algorithm will be used in each iteration of the dynamic MaaS-based ATP generator algorithms to periodically find the best ATPs for MaaS participants based on the rideshare opportunities that appear in the system.

| Algorithm 1: STATIC |
|---|
| 1  **Input:** Set of driver announcements $D^S \cup D^M$, set of passenger announcements $R^S \cup R^M$, $w_{d,r,\delta}$ |
| 2  **Output:** A weighted bipartite graph $G$, a matching vector $\boldsymbol{x}$ for all pairs of driver-passenger in $\bar{P}$ |
| 3  $P \leftarrow \{(d,r,\delta): d \in D^S \cup D^M, r \in R^S \cup R^M, \delta \epsilon E(d) \cup E(r)\}$ |
| 4  $\bar{P} \leftarrow \emptyset$ |
| 5  **for** $(d,r,\delta)$ in $P$: |
| 6      **if** $TF(d,r,\delta) = $ True, and $\Delta S(d,r) \geq 0$ **then**: |
| 7          $\bar{P} \leftarrow \bar{P} \cup \{(d,r,\delta)\}$ |
| 8      **end if** |
| 9  **end for** |
| 10  $\boldsymbol{w} \leftarrow$ Determine weight vector based on the objective function and $\bar{P}$ |
| 11  $G \leftarrow (D^S \cup D^M \cup R^S \cup R^M, \bar{P}, \boldsymbol{w})$ |
| 12  $\boldsymbol{z} \leftarrow$ Execute the generalized maximum-weight matching formulation proposed in Section 4.4 on $G$ |

### 4.6. Rolling Horizon Framework

Single-trip and MaaS-based announcements may enter the rideshare oriented MaaS system continuously at any time, thus making the problem dynamic. In this section, we introduce a *rolling horizon* approach to solve the on-demand tour-based ridesharing problem in real-time. The rolling horizon approach is motivated by the following concepts. First, the system is highly dynamic so that announcements enter and leave the system continuously. Second, the routes for MaaS-based announcements may be continuously re-considered at a later stage based on the other participants' announcements. Third, future driver and passenger requests are assumed to be unknown.

For running the rolling horizon-based model, we adopt periodic optimisation with a fixed time step (Najmi et al., 2017) which assumes that the rolling horizon algorithm is executed for a given set $\bar{T}$ of time steps, i.e.: $\bar{T} = \{0, h, 2h, 3h, \dots\}$. Also, we adopt an infinite look-ahead time horizon approach to slide the horizon with a predefined time step $h$, and then execute the STATIC algorithm to match or update the pairs for each time steps. Then, in each iteration of the rolling horizon, the matching problem presented in Section 4.4 is solved for the set of active announcements. At any time $t \in T$, an announcement is *active* if 1) it is for single-trip and its latest departure time is greater than the execution time of the STATIC algorithm, i.e. $q(p) \geq t$, or 2) it is for a MaaS-based participant and there is at least one activity in the activity list of the participant that its latest departure time is greater than the execution time of the STATIC algorithm, i.e. $\bar{t}_i^p \geq t$. If a single-trip announcement is labelled inactive if it is matched to another announcement or expired if $q(a) < t$ (Najmi et al., 2017). Also, a MaaS-based announcement is labelled inactive if the activity list is empty.

It is noteworthy that calculating all combinations of feasible pairs is computationally cumbersome because MaaS-based participants remarkably increase the number of pairs. While there is only one link (possible trip) for single-trip announcements, the MaaS-based participants have a network of possible trips which affects the number of the possible candidate rideshare links $\delta$. Therefore, the most important links for ridesharing plan should be selected for rideshare consideration. In line with this, in this paper, we only consider those candidate rideshare links $\delta$ for MaaS-based participants that are adjacent to the node that has accommodated the participant at the execution time of the STATIC algorithm. At any point of time, the model generates ATPs of participants for all the activities that have not been visited yet.

In each iteration of the rolling horizon framework, depending on the type of announcement, different strategies could be adopted. If the announcement is for a single trip, as in Najmi et al. (2017) which is recognised as *dynamic matching policy*, a matched pair $(d, r, \delta)$ can either be finalised, i.e. the match is accepted by the ridesharing system, or its finalisation can be delayed. Postponing the finalisation time has the advantageous of finding a better match for either the driver, the passenger or both of them in the future. For the pairs including MaaS-based announcements, owing to the uncertain routes for the participants, calculating the flexibility of the trips in the tour and thereby the latest departure time is complicated. Hence, we use As Soon As Possible (ASAP) policy for the MaaS-based participants. Under ASAP policy, any matched trip among announcements is finalised on its first occurrence within the rolling horizon algorithm. Thus, its finalisation condition for the ASAP policy is $z_{d,r,\delta} = 1$. Let denote $\bar{z}_{d,r,\delta}$ the finalised value of variable $z_{d,r,\delta}$, i.e. $\bar{z}_{d,r,\delta} = 1$ means that the match $(d, r, \delta)$ is accepted by the ridesharing system and $\bar{z}_{d,r,\delta} = 0$ otherwise which means that at least one of the participants in this matched pair is involved in another finalised match, thus these two announcements are not matched together. In contrast, for S&S, we use the As Late As Possible (ALAP) policy where matched trips $(d, r)$ are not finalized until the next time period exceeds the latest departure time of either driver or passenger, $q(d) < t + h$ or $q(r) < t + h$.

If a single-trip announcement is finalised, the respective requests for drive or ride exit the system and will not be considered at the next iterations. For MaaS-based announcements, the destination (activity) of the finalized rideshare link $\delta$ omits the activity list of the passenger; so, the passenger's ATP for the next iteration does not include the activity. The pseudo code of the rolling horizon algorithm is presented in Algorithm 2 and henceforth referred to as the ROLLING HORIZON algorithm.

| | Algorithm 2: ROLLING HORIZON |
|---|---|
| 1 | **Input:** Announcements sets $D^S, D^M, R^S$ and $R^M$, objective function type |
| 2 | **Output:** A finalised vector $\bar{\mathbf{z}} = [\bar{z}_{dr\delta}]$ |
| 3 | **for** $t \in \{0, h, 2h, 3h, ...\}$: |
| 4 | $\quad D_t^S \leftarrow \{p \in D^S: q(p) \geq t, a(p) \leq t\}$ |
| 5 | $\quad R_t^S \leftarrow \{p \in R^S: q(p) \geq t, a(p) \leq t\}$ |
| 6 | $\quad D_t^M \leftarrow \{p \in D^M: |V(p)| > 0\}$ |
| 7 | $\quad R_t^M \leftarrow \{p \in R^M: |V(p)| > 0\}$ |
| 8 | $\quad G \leftarrow$ STATIC $(D_t^S, R_t^S, D_t^M, R_t^M,$ objective function$)$ |
| 9 | $\quad \mathbf{z} \leftarrow$ Execute the maximum-weight bipartite matching algorithm on $G$ |
| 10 | $\quad$ **for** $(d, r, \delta) \in \bar{P}_t : x_{dr\delta} = 1$: |
| 11 | $\quad\quad$ **if** MatchingType = S&S **then:** |
| 12 | $\quad\quad\quad$ **if** $\min[q(d), q(r)] < t + h$: |
| 13 | $\quad\quad\quad\quad \bar{z}_{dr\delta} \leftarrow 1$ |
| 14 | $\quad\quad\quad\quad D_t^S \leftarrow D_t^S \setminus \{d\}$ |
| 15 | $\quad\quad\quad\quad R_t^S \leftarrow R_t^S \setminus \{r\}$ |
| 16 | $\quad\quad\quad$ **end if** |
| 17 | $\quad\quad$ **else:** |
| 18 | $\quad\quad\quad \bar{z}_{dr\delta} \leftarrow 1$ |
| 19 | $\quad\quad\quad$ **if** $d \in D_t^S$ **then:** |
| 20 | $\quad\quad\quad\quad D_t^S \leftarrow D_t^S \setminus \{d\}$ |
| 21 | $\quad\quad\quad$ **else:** |
| 22 | $\quad\quad\quad\quad D_t^M \leftarrow V(p) \setminus \{j(\delta)\}$ |
| 23 | $\quad\quad\quad$ **end if** |
| 24 | $\quad\quad\quad$ **if** $r \in R_t^S$ **then:** |
| 25 | $\quad\quad\quad\quad R_t^S \leftarrow R_t^S \setminus \{r\}$ |
| 26 | $\quad\quad\quad$ **else:** |
| 27 | $\quad\quad\quad\quad R_t^M \leftarrow R_t^M \setminus \{j(\delta)\}$ |
| 28 | $\quad\quad\quad$ **end if** |
| 29 | $\quad\quad$ **end if** |
| 30 | $\quad$ **end for** |
| 31 | $\quad$ **for** $p \epsilon (D_t^S \cup R_t^S)$: |
| 32 | $\quad\quad$ **if** $q(p) < t + h$ **then:** |
| 33 | $\quad\quad\quad$ remove the announcement from either $D_t^S$ or $R_t^S$ |
| 34 | $\quad\quad$ **end if** |
| 35 | $\quad$ **end for** |
| 36 | $\quad$ **for** $p \epsilon (D_t^M \cup R_t^M)$: |
| 37 | $\quad\quad$ **if** the departure time of the first trip in the ATP is due **and** the first trip in the ATP is the same as the first trip in the ATP for previous iteration **then:** |
| 38 | $\quad\quad\quad$ remove the activity node of the first trip in the ATP from the activity list of $p$ |
| 39 | $\quad\quad$ **end if** |
| 40 | $\quad\quad$ **if** there is no feasible ATP for $p$ **then:** |
| 41 | $\quad\quad\quad$ remove the announcement from either $D_t^M$ or $R_t^M$ |
| 42 | $\quad\quad$ **end if** |
| 43 | $\quad$ **end for** |
| 44 | **end for** |

## 5. Numerical Experiments

In this section, we evaluate the performance of the proposed rideshare oriented MaaS model. The pre-processing stage is computationally extensive. In this paper, we solve it on a medium-size experiments for illustration purpose.

### 5.1. Data and Simulation

To test the performance of the proposed rideshare system, we use tempo-spatial data from the inner and eastern suburbs of Sydney, Australia. Sydney is the most populated city of Australia and capital of the state of New South Wales. As it is depicted in Fig. 8, we consider pools of 20 shopping centres, 20 service centres, and 10 educational centres to be chosen in the experiments. Furthermore, we assume that home and work locations are uniformly distributed over the experiment area.

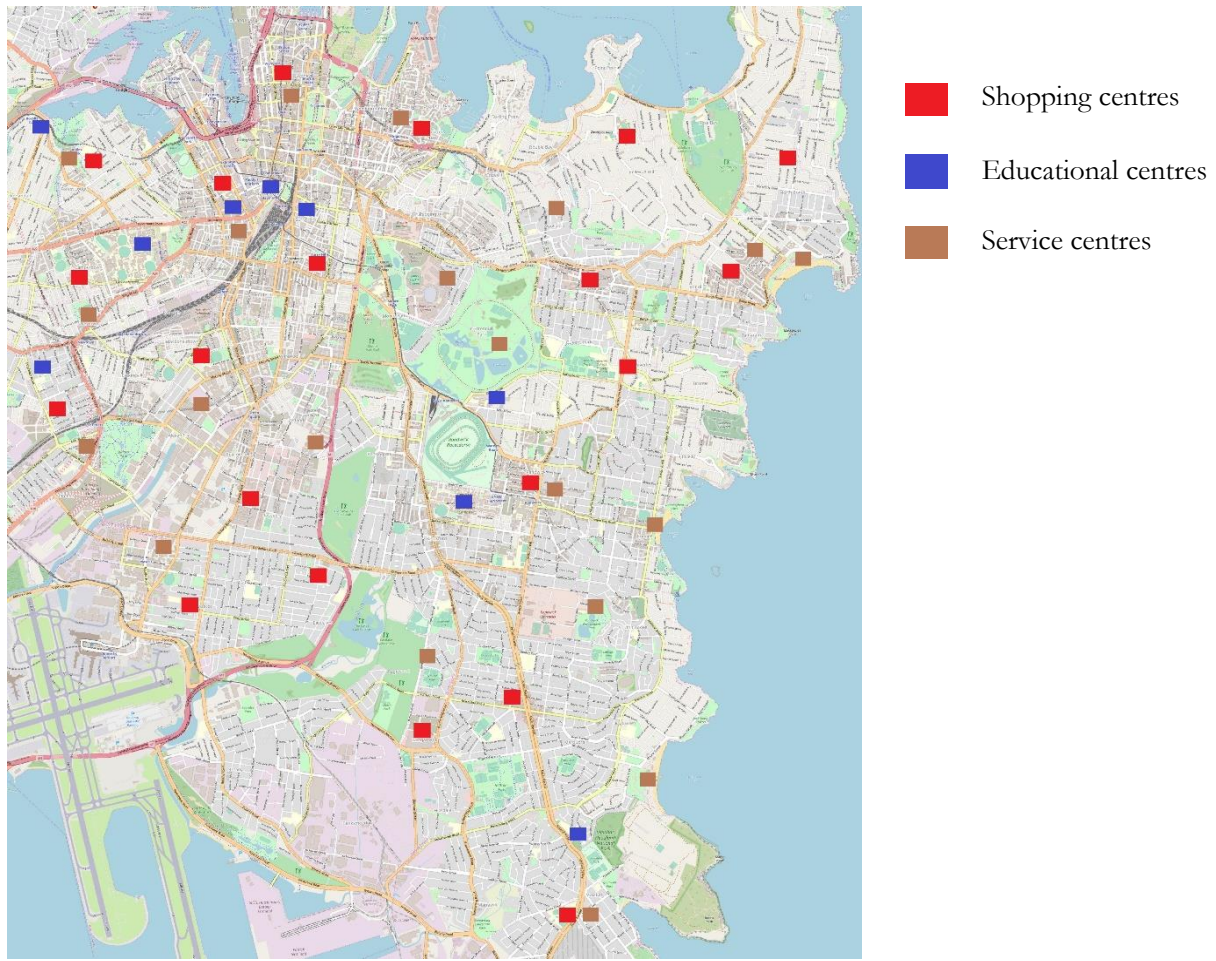

Figure 8 Simulation area

There are two different request types of MaaS and single-trip in the system. For MaaS requests, there are some attributes that are randomly generated for each participant and are kept constant throughout the experiments. The attributes include home, work and educational locations, as well as the trip purposes that should be met over the period. Also, each participant has a minimum and maximum number of trips of 2 and 5 over the scheduling period of 7:00am to 18:00pm. The parameters in Table 1 are used to generate the synthesised datasets. Three random replications of participants with randomly selected attributes are generated to be able to accurately compare and assess the performance of different variants of the model. To generate the random streams, we firstly generate the number and types of activities to visit. We consider four activity types of work, shopping, service and education each of which have a chance to be included in activity list of each participant.

**Table 1** Parameters used to generate synthetic population.

| Parameters | Values |
|---|---|
| Number of shopping center locations | 20 |
| Number of service center locations | 20 |
| Number of educational locations | 10 |
| Commuting Probability | 0.7 |
| Going to a service center probability | 0.6 |
| Going to a shopping center probability | 0.8 |
| Going to an educational center probability | 0.5 |
| Travel-time coefficient | 0.2 |
| Time spent at home before leaving | Randomly from [0,350] |
| Work duration | Randomly from [300,540] |
| Service duration | Randomly from [15,120] |
| Shopping duration | Randomly from [15,120] |
| Education duration | Randomly from [240,360] |

After determining the activity types and their sequences, the activity locations are determined. We differentiate between different activities to be met. In these experiments, the home locations for each participant is determined first, and then the location of fixed activities such as work and school are generated by giving the higher weights to the nodes far from their home. This is intentionally done to increase the chance of finding the matches when solving the model. For this purpose, a simple logit model is used with the travel-time coefficient parameter provided in Table 1. After assigning the home and fixed activity locations, the flexible activity locations are generated for each participant. For each of the participants, we randomly select 3 alternative destinations for each flexible activity type if the activity is in their activity list to visit. We assume that either the work trip or the educational trip can be included in an activity travel pattern, if any. The time window for the departure time and the duration of activities are randomly selected using a uniform distribution from the ranges provided in Table 1. Furthermore, we use a relatively large time window for the departure time of the flexible activities to allow rescheduling the participants' travel patterns easily.

For single-trip based requests, we use the same location pools, logit model, and activity durations as for MaaS requests; however, the generated trip requests for the experiments are fixed and for a single trip for each participant. The departure time for the requests are randomly selected from the scheduling timeframe. For trip-based requests, we fixed trip flexibility rather than allowing it to be a function of travel time. Let $b$ and $c$ be departure time and travel time of a trip generated. We determine its earliest departure time and its latest arrival time as $\underline{t}_i^p = b - 40$ and $l = b + c + 40$ (in minutes), respectively. Finally, we randomly generate announcement times using a uniform distribution with parameters ($b$ - 90, $b$). It should be noted that, in this paper, compared to single-trip requests, we assume that all MaaS requests are received before the start of the scheduling period (day). A period of $h = 20$ min (time step) is used for the rolling horizon algorithm which results in a total of 33 time periods ($|T| = 33$). We assume full compliance of the participants, i.e. if a ride is finalised, then the corresponding driver and passenger are immediately notified and accept the finalised match. Furthermore, two participation scenarios of 2000 and 3000 participants are considered. We consider %30 and %70 of participants as MaaS-based and Single-trip based, respectively.

To evaluate the performance of the proposed MaaS model, three variants of the model are used which are abbreviated as follows: DSM (Dynamic Single-trip based Matching), SSM (Static Single-trip based Matching), and DMM (Dynamic MaaS-based Matching). DSM is the same as the models in Agatz et al. (2011) and Najmi et al. (2017) and is discussed in this section for comparison purposes (as the benchmark). In DSM, the classic dynamic matching formulation is used to find the matching rate. In this variant, the optimum ATPs for individuals are considered as the travel demand and fixed. To illustrate more, an optimum tour of length 3 is considered as 3 single trips with fixed origins and destinations. To obtain the optimum pattern, we run ATP generator only once to find the best ATPs for travellers if they ignore the behaviour of other participants. To assess the performance of dynamic models, it is a standard practice to compare its output with its static counterpart, wherein all announcements are known prior to the start of the day to assess the performance. Therefore, in SSM, all the settings are as the same as for DMM except that all announcements are known prior to the start of the day. In DMM, which is the core in this research, the activity lists are pre-defined but the ATPs are not fixed and should be dynamically optimised (rescheduled). It should be mentioned that the static version of DMM model is unsolvable, as it is discussed in Section 5.2.2 about the model complexity; so, it is ignored here. Other than running the three variants of the model, we run a greedy algorithm over the requests in the synthesised datasets for comparison purposes. This is helpful in gaining some understanding of the value of the optimization-based approaches in rideshare matching.

It should be highlighted that although the proposed model considers the flexible role of participants, in the scenario analysis, we assume that the role of all participants is determined and fixed, as drivers or passengers. We assume that if the participants are under driver category, they have access to only driving links; on the other hands, the passengers have access to all the outbound links except driving links. We also assume that the cost on all the links are proportional to the travelled distances. All algorithms of the proposed model are implemented in Python 2.7 on a machine with 16 Gb of RAM with a processor of i7-4770. The routing problem is solved by calling CPLEX.

## 5.2. Computational results

In this section, we first present the evaluation criteria that we use for performance assessment and then the simulation results of the variants explained before are discussed.

### 5.2.1 Performance evaluation

In this paper, we consider the following frequently used performance measures in the literature to evaluate the performance of the proposed rideshare oriented MaaS model; nonetheless, the performance measures may have different weights across different decision makers.

1. Matching rate (MR): the total number of matched driver and passenger announcements divided by the total number of trip announcements;
2. Average total vehicle-kilometres savings (AKS): total kilometres saved as a result of matching algorithms versus the scenario in which all individual trips are performed; and

### 5.2.2 Dynamic Problem Benchmark

To conduct the dynamic problem analysis, the model variants introduced in Section 5.1 are used in conjunction with two objective functions of maximisation of number of matches (NM) and maximisation of distance savings (DS). Despite we have discussed two different objective functions here, the detailed comparison between their results is not under the scope of this study. They are discussed in Najmi et al. (2017) in detail. Our main objective in this section is to show the suitability of including MaaS-based requests (with ATP generator) in the traditional rideshare systems. The objective function NM is discussed here only to obtain the extreme value for MR and to show the existing opportunity for future research to improve the matching model to enhance the quality of the solutions.

Table 2 summarises the results of running different variants with both objective functions for matching. The table depicts clearly that the Greedy approach is significantly outperformed by other variants in terms of MR and AKS based on the participation rates. While the greedy algorithm seems reasonable, it does not result in proper output in practice. Despite the fact that the greedy algorithm may generate acceptable individual matches, it does not perform well at the system level, not yielding proper matching rates nor average cost savings.

Among the static variants, NM-SSM (in this paper, for instance, NM-SSM refers to a NM objective function that is implemented for the SSM variant) and DS-SSM, respectively, determine the extreme values for MR and AKS in the single-trip based variants. The gap between the SSM and DSM variants demonstrates the potential improvement when prior higher level of information becomes available to participants before finalising matches. It can be seen that there are significant potentials for Improving DSM outputs if more information about the requests is acquired. Nonetheless, the results reveal that entering MaaS requests in the system is another option to upgrade the system performance (see the results for DMM variant).

Comparing to DSM, DMM significantly improves the quality of the objective functions in terms of both MR and AKS. While the improvement in MR for DS is about 9%, the value is 17.2% for NM. Furthermore, the improvement in AKS for DS is about 24% while the corresponding value for NM is more than 65%. Although the DMM variant shows the considerable impacts of the activity generator in MaaS on the NM, still the objective function has a bad performance in term of AKS. Another finding of the results in the table shows the importance of the participation rate. Similar to the results in Stiglic et al. (2016) and Najmi et al. (2017), the MR and AKS for the scenario with higher spatial density of participants is significantly higher which shows the important of participation density in ridesharing systems. However, the improvement is marginally significant for DS in term of AKS.

**Table 2.** Performance measures for the dynamic problem benchmark

| Scenario | Variant | Model | Matching objective function | | | |
|---|---|---|---|---|---|---|
| | | | NM | | DS | |
| | | | MR | AKS | MR | AKS |
| Scenario 1 | V1 | Greedy | - | - | 12.23% | 1.60% |
| | V2 | DSM | 22.18% | 2.10% | 18.44% | 5.22% |
| | V3 | SSM | 25.18% | 3.45% | 21.16% | 6.49% |
| | V4 | DMM | 26.00% | 3.43% | 19.78% | 6.38% |
| | | Improvement of DMM over DSM | 17.22% | 63.33% | 7.27% | 22.22% |
| Scenario 2 | V1 | Greedy | - | - | 12.28% | 1.66% |
| | V2 | DSM | 24.69% | 2.17% | 21.55% | 6.61% |
| | V3 | SSM | 27.31% | 3.54% | 24.32% | 8.10% |
| | V4 | DMM | 28.94% | 3.60% | 23.53% | 8.26% |
| | | Improvement of DMM over DSM | 17.21% | 65.90% | 9.19% | 24.96% |

Table 3 summarises the detailed statistics and performance of DSM and DMM with higher participation rate scenario. In the table, the ATP-based trips and MaaS-based requests incorporate both the requests for drive and ride in the system. For DSM, we do not use MaaS-based terminology as it does not equipped with ATP rescheduling to modify the ATPs based on others' schedules; instead, it only generates a single ATP for each tour-based participant at the start of the day. So, we have simply called it ATP-based trips which denote the trips that are generated in the ATP generation step. Comparing the ATP-based trips and single-trip statistics for DSM variant reveals the noticeably bad performance of the ATP-based trips. Although the ATP-based trips comprise 44.9% (1956 over 4356) of the total number of trips, it only accounts for 32.38% and 17.58% of the saving and success rate contributions. Also, the average saving for this category is considerable bad (0.45 km vs 0.76 km). The reason for the bad performance is the existence of many trips with short lengths (in comparison with Single-trips) in the synthesised data for DSM. The short trips are generated because the ATP generator selfishly looks for the route with the least cost for each participant. In Agatz et al. (2011) and Najmi et al. (2017), the authors show the low chance of short trips in finding matches in rideshare systems.

Comparing the MaaS-based and single-trip statistics for DMM variant highlights the following issues: First, having the opportunity to choose among multiple destinations (ATP generation) plays a key role in changing the utility of participants. While the total number of trips for MaaS-based requests accounts for 44.9% of the total number of trips, they incorporate 60.48% of the total saving in the system. Second, the comparison of success rate for MaaS-based and single trip requests reveals that 39.83% of MaaS-based participants are matched at least once which is much higher than the corresponding value (about 13%) for single-trip based. These show the remarkable effects of accommodating ATP generation with rescheduling in MaaS system in enhancing the performance of the system by increasing the success rate and distance saving measures. Furthermore, the MaaS-based participants have obtained higher average distance saving (almost double) compared to single-trip based participants which is opposed to the the ATP-based trips in DSM.

Overall, by comparing the performance DSM and DMM in Table 3, we can conclude that the incorporation of ATP-based trips in DSM is beneficial for single-trip participants because this raises the density of the participants which results in better performance for single-trip participants. Nonetheless, the condition is quite different for MaaS-based participants in DMM. The rescheduling of ATP generator in DMM significantly increases the contribution of MaaS-based participants (and accordingly decreases the contribution of participants with Single-trip announcements) in the system performance. An important conclusion is that while the low chance of ATP-based trips (in comparison with the single trip announcements) may encourage the participants to leave the system, the higher chance to find a match and the higher average saving rate are incentives for single-based participants to shift to an MaaS-based stream which is armed with rescheduling.

**Table 3.** Detailed performance of the rideshare model

| Model | Requests | Total No. of requests | Total No. of trips | Total original distances (km) | Distance saving contribution (km) | Saving contribution (%) | Success rate contribution (%) | Saving contribution per trip (km) |
|---|---|---|---|---|---|---|---|---|
| DSM | ATP-based trips | 600 | 1956 | 7279.52 | 875.53 | 32.38% | 17.58% | 0.45 |
|  | Single trips - Riders | 1200 | 1200 | 6634.2 | 901.25 | 33.33% | 24.41% | 0.75 |
|  | Single trips - Drivers | 1200 | 1200 | 6537.23 | 926.84 | 34.28% | 25.13% | 0.77 |
|  | Sum | 3000 | 4356 | 20450.96 | 2703.62 | 100 |  |  |
| DMM | MaaS-based requests | 600 | 1956 | 7279.52 | 2043.35 | 60.48% | 39.83% | 1.04 |
|  | Single trips - Riders | 1200 | 1200 | 6634.2 | 687.06 | 20.34% | 13.25% | 0.57 |
|  | Single trips - Drivers | 1200 | 1200 | 6537.23 | 648.1 | 19.18% | 13.92% | 0.54 |
|  | Sum | 3000 | 4356 | 20450.96 | 3378.5 | 100 |  |  |

Fig. 9 shows the matching rate patterns of MaaS-based requests versus their total number of matched trips which are categorised based on their original distances from home to fixed destinations (work or education purposes) (henceforth referred to as the home-based tour gyration). The tour gyration is important as it can significantly limit the choice of routes for the participants. As it can be perceived from the figure, the MaaS-based requests with the tour gyration of ≤2km and 2km<x≤5km are with 51% and 42% probability of failure to find even one match in their tour. The success rates for the tour gyration of 5km<x≤8km and ≥8km are about 13% and 9.5% respectively which shows the significant role of tour gyration in finding a match. The MaaS-based requests with the shorter tour gyration (less than 5 km) have on average 42% chance to find exactly one match; while the value drops to around 10% and 0 on average for finding 2 and 3 matches, respectively. Interestingly, the MaaS-based requests with longer tour gyration (greater than 5 km) have more than 50% chance to find at least 2 matches. For the tour gyration of greater than 8 km, there are the chances of 25% and 8%, respectively, to find at least 3 and 4 matches in the tour. These shows the critical role of tour gyration in the MaaS-based ATP generator which can encourage people to participate in MaaS system. For example, the government authorities can use the outstanding results of the tours with high radius of gyration to incentivise people who commute long distances on a daily basis to join the MaaS system.

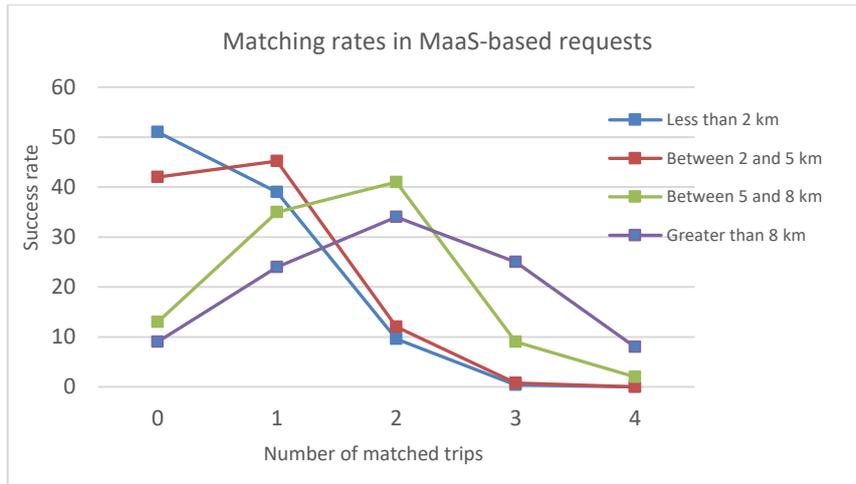

Figure 9 Potential of MaaS-based planning in ridesharing context for different tour gyration

A closer inspection of the profile of distance saving based on their original distances for different types of requests shows that the higher distance of the original trip length is, the more the saving may be obtained.

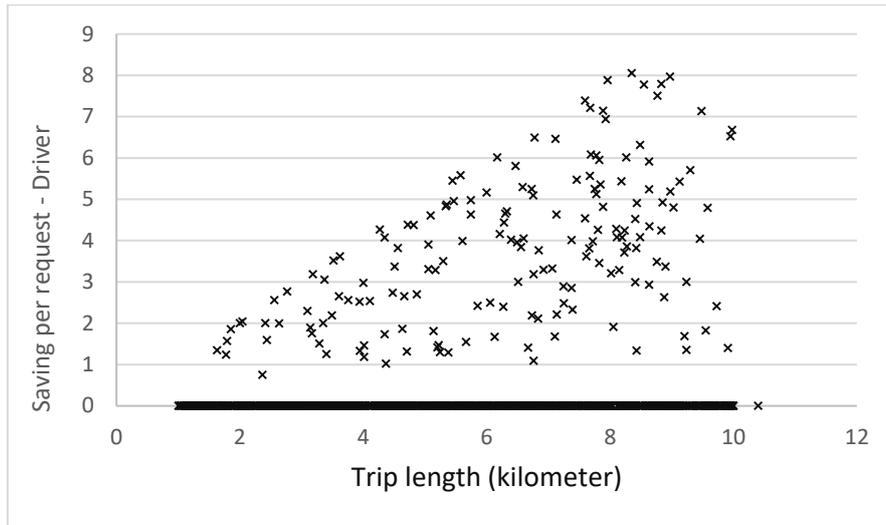

a) Drivers' role in saving

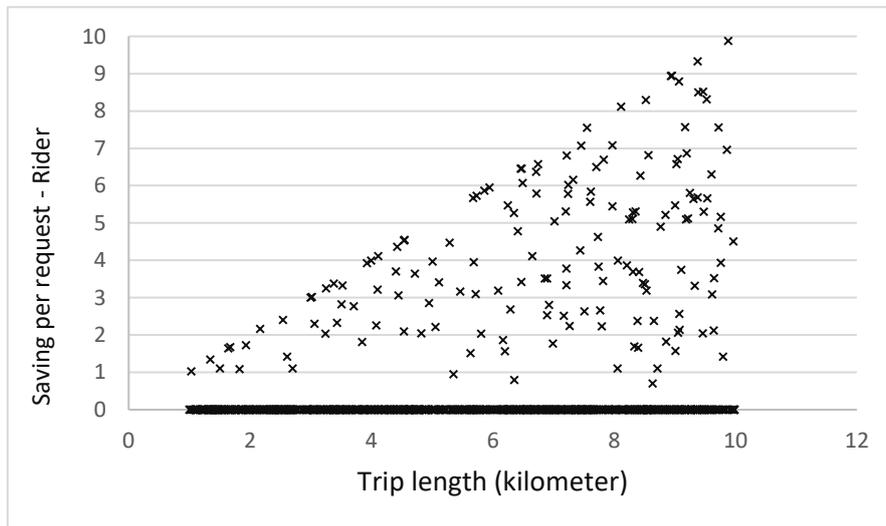

b) Drivers' role in saving

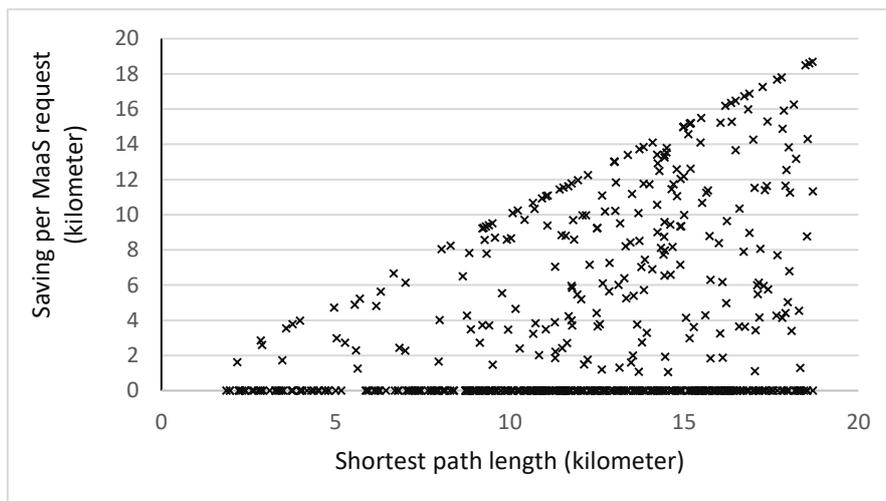

c) MaaS-based requests role in saving

Figure 10 Saving distances vs original route lengths

Next, we investigate the probability of finding a match based on participant's trip length. Fig. 11 breaks down trip announcements based on the participants' individual trip distance to investigate their success rate in finding a match. As shown in the figure, the probability of finding a match for drivers steadily increases with an increase in the trip length. Longer trips, in both the DSM an DMM cases, increase the probability of finding a compatible passenger both en-route and detour which can then result in having more potential distance savings. The MaaS-based requests positively affect the success probability of the drivers in all categories. For passenger trips, we find that the probability of finding a match for DSM an DMM has different patterns. While the MaaS-based announcements raise the success probability of passengers with trips of length between 2 and 8 kilometres, the increase is more significant for longer trips (>8). Interestingly, the MaaS-based requests declines the success rate for the short trips of less than 2 kilometres. This is mainly due to the rescheduling opportunity in the DMM variant where some participants choose flexible destinations with farther distance to increase their matching success rate.

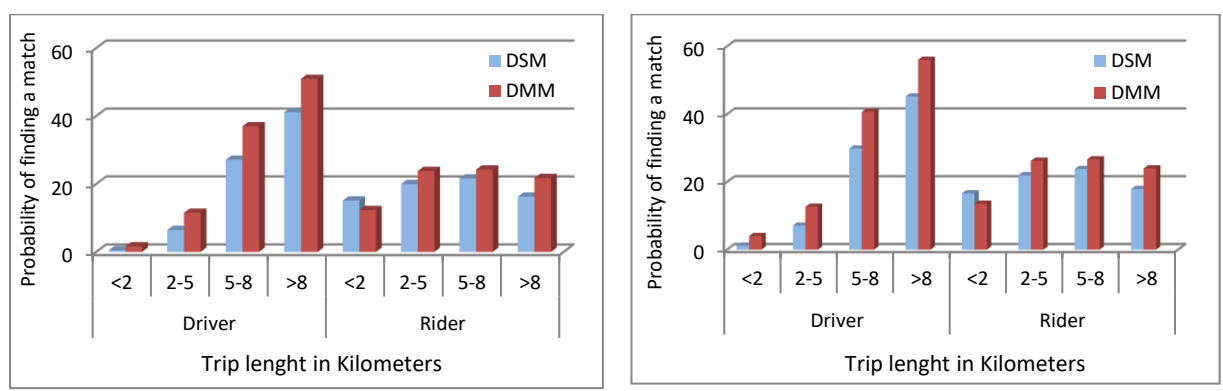

a) Scenario 1 (low participation rate)  b) Scenario 2 (medium participation rate)

Figure 11 Probability of finding a match vs. participants' individual trip distance.

The discussion is this section revealed the considerable role of ATP generation in the performance of shared mobility systems.

## 6. Conclusion

*Summary of findings:*

In this paper, we proposed a novel MaaS-based ATP generator to dynamically reschedule travelling plan of people. We also extended the ridesharing literature by incorporating the ATPs in rideshare searches for matching. In line with these, the MaaS-based ATP generator and rideshare model are synchronised to enhance mobility of people. The models are on-demand which interactively influence each other. This allowed exploring the role of the ridesharing concept on the ATPs of individuals and vice versa. We found that the determining the activity tour by platform (MaaS-based ATPs) can substantially improve the performance of the ridesharing system.

The presence of MaaS-based announcements in the system has a remarkable impact on the performance of the rideshare systems as not only they increase the success rates of finding a match in the system but also do increase the attractiveness of MaaS mode for the participants.

Also, the results showed that the MaaS-based participants in the system, despite their fewer number of participants, incorporate a major portion of the successful participants in finding a match. Furthermore, the participants incorporate the biggest share of total saving contribution in the system. Other than these, we found that the MaaS-based participants have obtained higher average distance saving compared to single-trip based participants. The significant positive effects of synchronising the rideshare model and MaaS-based ATP generator may encourage government and private companies to introduce incentive mechanisms to increase the number of participants in MaaS-based ATP planning program. The higher chance to find a match and the higher saving rate are incentives for single-based participants to shift to MaaS-based stream. Furthermore, we showed the key role of tour gyration on the performance of MaaS system. The outstanding performance of participants with high tour gyration is another incentive for people with distant activity locations to join MaaS systems.

*Future direction:*

This paper aimed at introducing a new definition for MaaS and proposing a novel model for its integration with rideshare systems. This opens an avenue to expand the opportunities to increase the mobility in societies. First, due to the complexity of the model, we did not consider the flexibility of ATPs and their included trips; so, we naïvely matched trips in the MaaS-based ATPs if they have found a match. However, a better solution might be obtained if a wiser finalization matching policy is used. Due to the tour-based nature of MaaS-based requests, calculating the time and departure time flexibility for all the MaaS participants in a short period of time would be complicated. Therefore, an extension of this paper would be proposing a rolling horizon based matching policy to postpone the finalization of matches hoping to find better matching counterparts. Second, the carsharing which is another shared mobility component can be added to the MaaS-based mobility model by modifying the pre-processing procedure. Then, the MaaS platform iteratively updates the ATPs of travellers as well as the real-time availability of the shared cars in the system. Nonetheless, the optimal assignment of the shared cars in ATPs generation process, which is different than serving the first in first served policy for the shared cars, is a complicated problem that should be addressed. Third, the problem is computationally extensive and needs extensive pre-processing which makes solving the problem in real-time complicated. Therefore, another potential future research is to solve the rideshare oriented MaaS model for large-scale examples. Fourth, we assumed that the participants accept the ATPs and matched trips that are generated by the system. In other words, we developed an automated system that establish and reschedule ATPs and rideshare matches with minimal input from participants and improve system-wide performance measures. Nonetheless, system-level considerations may not provide the maximum benefit to each individual participant. Therefore, a potential future research would be investigating interactive behaviour of participants in the MaaS system and its influences on the performance of the system.